\documentstyle[aps,multicol,epsfig]{revtex}
\begin{document}
\draft

\title{Stripes in La$_2$NiO$_{4.17}$: charge and spin ordering
according to NMR.}

\author{I.M. Abu-Shiekah, O.O. Bernal\cite{OOB}, H.B. Brom, M.L. de
Kok, A.A. Menovsky, J.T.  Witteveen}

\address{Kamerlingh Onnes Laboratory, Leiden University,\\
         P.O.Box 9506, 2300 RA Leiden, The Netherlands}

\author{J. Zaanen}

\address{Instituut Lorentz for Theoretical Physics, Leiden
University,\\
P.O. Box 9506, 2300 RA Leiden, The Netherlands}

\date{\today}

\maketitle

\begin{abstract}
Charge and spin ordering in La$_2$NiO$_{4.17}$ has been studied by $^{139}$La
NMR using the field and temperature dependence of the
linewidth and relaxation rates. The stripes are site centered and
the domain spins freeze at $\sim 180$~K, followed by a second transition at
$\sim 50$~K where the spins on the domain walls freeze. Although consistent
with the neutron measurements, the results are quantitatively different from
the NMR results in a nickelate with quenched Sr disorder.
\end{abstract}


\begin{multicols}{2}
\settowidth{\columnwidth}{aaaaaaaaaaaaaaaaaaaaaaaaaaaaaaaaaaaaaaaaaaaaaaaaa}

Evidence is accumulating that the electron systems in doped Mott-Hubbard
insulators exhibit quite complex ordering phenomena\cite{Emery93}. In
two dimensional (2D) systems this takes the form of stripe phases where
the excess charges bind to antiphase boundaries in the N\'eel
state \cite{Zaanen89}. For obvious reasons the stripe phases in the
high $T_c$ cuprates \cite{Tranquada95} attract much attention, but they
also occur elsewhere: they were actually discovered in the nearly
isostructural doped La$_2$NiO$_{4}$ \cite{Chen93,Tranquada94}.
It is believed that quantum fluctuations are of less importance in
this system \cite{Littlewood94} and they might represent a more classical
version of the cuprate stripes. The microscopic characterization of these
stripes has been leaning heavily on neutron-\cite{Tranquada94,Lee97} and
electron diffraction \cite{Chen93} work. Here we will demonstrate that the
information obtained by NMR confirms the picture suggested by recent neutron
diffraction results \cite{Lee97}, adding to it a microscopic interpretation
of the spin system in the stripe phase, which turns out to be quite unusual.

Below we analyze the field and temperature dependence  of the
$^{139}$La linewidth and relaxation rates for La$_2$NiO$_{4+\delta}$
with $\delta= 0.17$. $^{139}$La has a nuclear spin $I=7/2$, which makes
NMR sensitive to both charge and spin, and allows the study of charge and
spin order and also the dynamics at time scales longer than $10^{-7}\, {\rm
s^{-1}}$. Our work is complementary to a recent NMR  study on
La$_{5/3}$Sr$_{1/3}$NiO$_4$ \cite{Yoshinary98}. The difference is that, due
to the oxygen ordering, the oxygen doped system is a much cleaner system, and
the  puzzling effects of quenched disorder in the Sr doped system causes
marked differences, obscuring the picture we find.

Neutron and susceptibility measurements \cite{Lee97} have revealed that
at $\simeq 240$K charge orders in a hexatic fluid- or glassy state, while at
$\simeq 190$K a first magnetic transition occurs accompanied by an improving
order in the charge sector, followed by a second transition at $\simeq 50$~K.
Our data reveal that for $T <180$~K the stripes are well localized (width of
order of a lattice constant) and precisely {\em site centered}, while the
first magnetic transition corresponds with a freezing of the $S=1$ spins in
the magnetic domains.  As pointed out by Zaanen and
Littlewood \cite{Littlewood94}, in the case of nickelates the charges
condensing on the  walls carry in addition a $S=1/2$ spin freedom, defining a
spin system living on the domain walls, which is for a precise site ordering
decoupled from the $S=1$ spin system. Our date reveal that this domain wall
spin system remains in a disordered state down to $50$ K, where a transition
follows into a fully static state. Fig.~1 summarizes the unusual nature of
the stripe state spin system.\\
\begin{figure}[htb]
\begin{center}
\leavevmode
\epsfig{figure=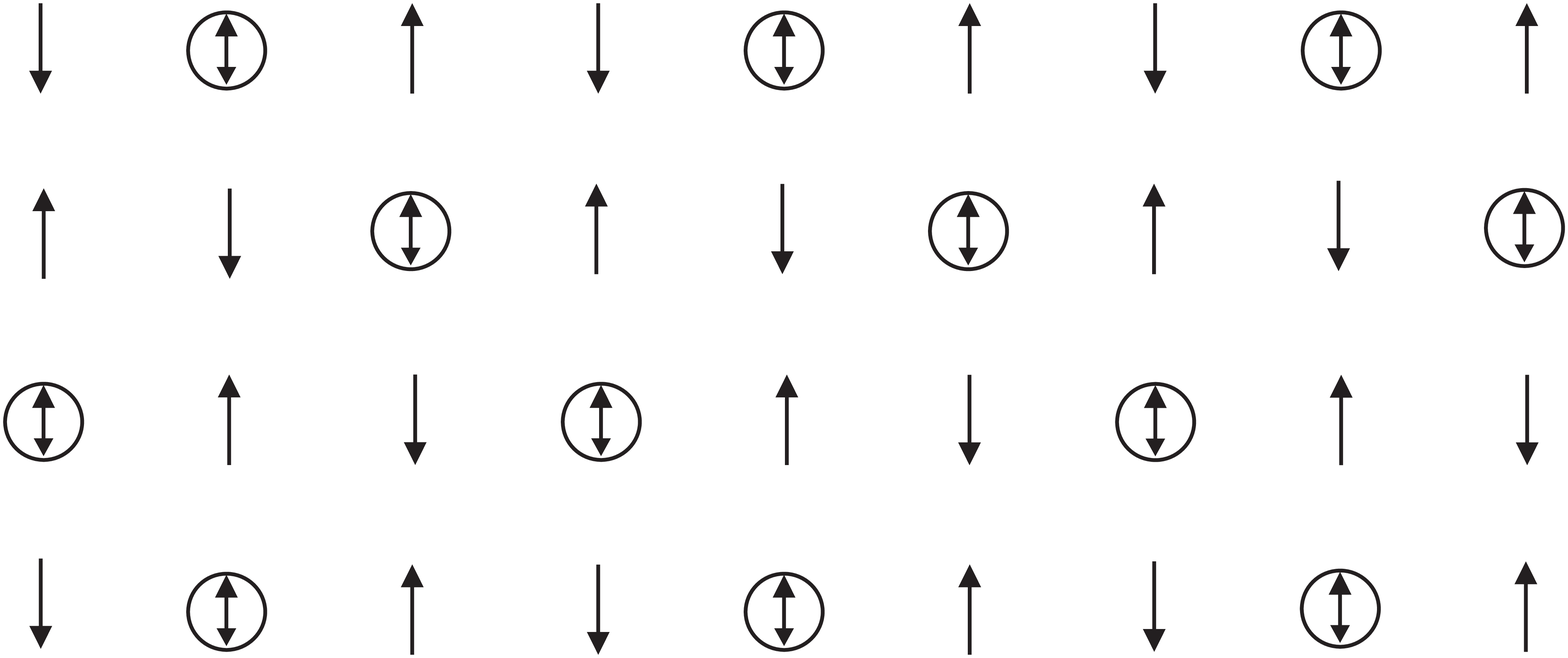,width=7cm,angle=0}
\end{center}
\caption{The spin system in the nickelates stripe phase at temperatures 50 K
$<$ T $<$ 190 K.  The $S=1$ spins in the magnetic domain exhibit a 2D order,
while the $S=1/2$ spins associated with the Ni$^{3+}$ `holes' living on the
domain walls remain disordered down to 50 K where 3D spin freezing sets in.
Notice that if the stripes are exactly Ni-site centered, the $S=1/2$ spin
subsystem is at least on the classical level completely decoupled from the
domain spin system.}
\label{f1}
\end{figure}
The measurements were performed on two single crystals from different batches
that were prepared under atmospheric condition in a mirror oven at 1100~K
\cite{Bernal97}.  Slices from both samples were analyzed by microprobe
techniques and showed oxygen gradients; on the average the oxygen contents
were found to be the same.  Samples for the measurements were cut from those
parts that had a homogenous oxygen content and had a typical weight of 10 mg.
Thermogravity (TGA) analysis of the oxygen concentration gave $\delta$ =
0.17.  For $\delta \geq 2/15$ the interstitials order three dimensionally
creating a large unit cell~\cite{Rodriguez91,Tranquada97}. The oxygen order
induces a tilt pattern of the NiO$_6$ octahedra that is observed in neutron
measurements \cite{Tranquada94}.  \\
Lineprofiles and relaxation data were measured in 9.4~T and 4.7~T and
typical results are presented in Figs.~2, 3, and 4.
\begin{figure}[htb]
\begin{center}
\leavevmode
\epsfig{figure=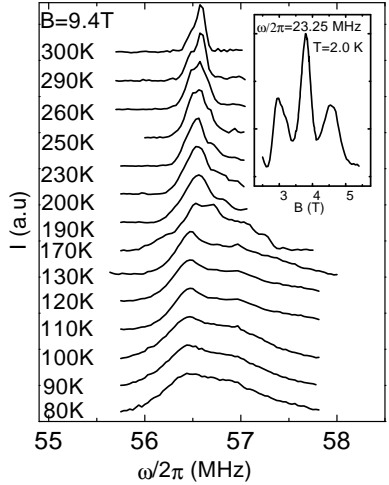,height=7cm,angle=0}
\end{center}
\caption{$^{139}$La NMR profiles in 9.4~T as a function of $T$.  Both
frequency and field sweeps were used.  Below 200~K the line is seen to split
into two lines (line A with a typical width of .25 MHz and a wider line B,
see Fig.3a for an example of the decomposition). The full field profile at
$\omega/2\pi= $~23 MHz at 2~K shows the location and the widths of three
quadrupolar transitions.}
\label{f2}
\end{figure}
\begin{figure}[htb]
\begin{center}
\leavevmode
\epsfig{figure=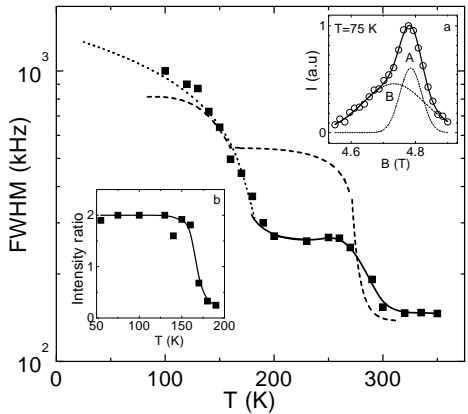,width=6.5cm,angle=0}
\end{center}
\caption{The $T$-dependence of the full width at half height of the total
resonance line in 9.4~T and 4.7~T (dashed) in a typical $T$ run.  Inset (b)
shows the intensity ratio of line B to A (an example of the decomposition of
a (field sweep) spectrum in two Gaussians is given in inset (a); Dashed and
drawn lines are guides to the eye; the dotted line below 200~K is a fit,
discussed in the text.}
\label{f3}
\end{figure}
\begin{figure}[htb]
\begin{center}
\leavevmode
\epsfig{figure=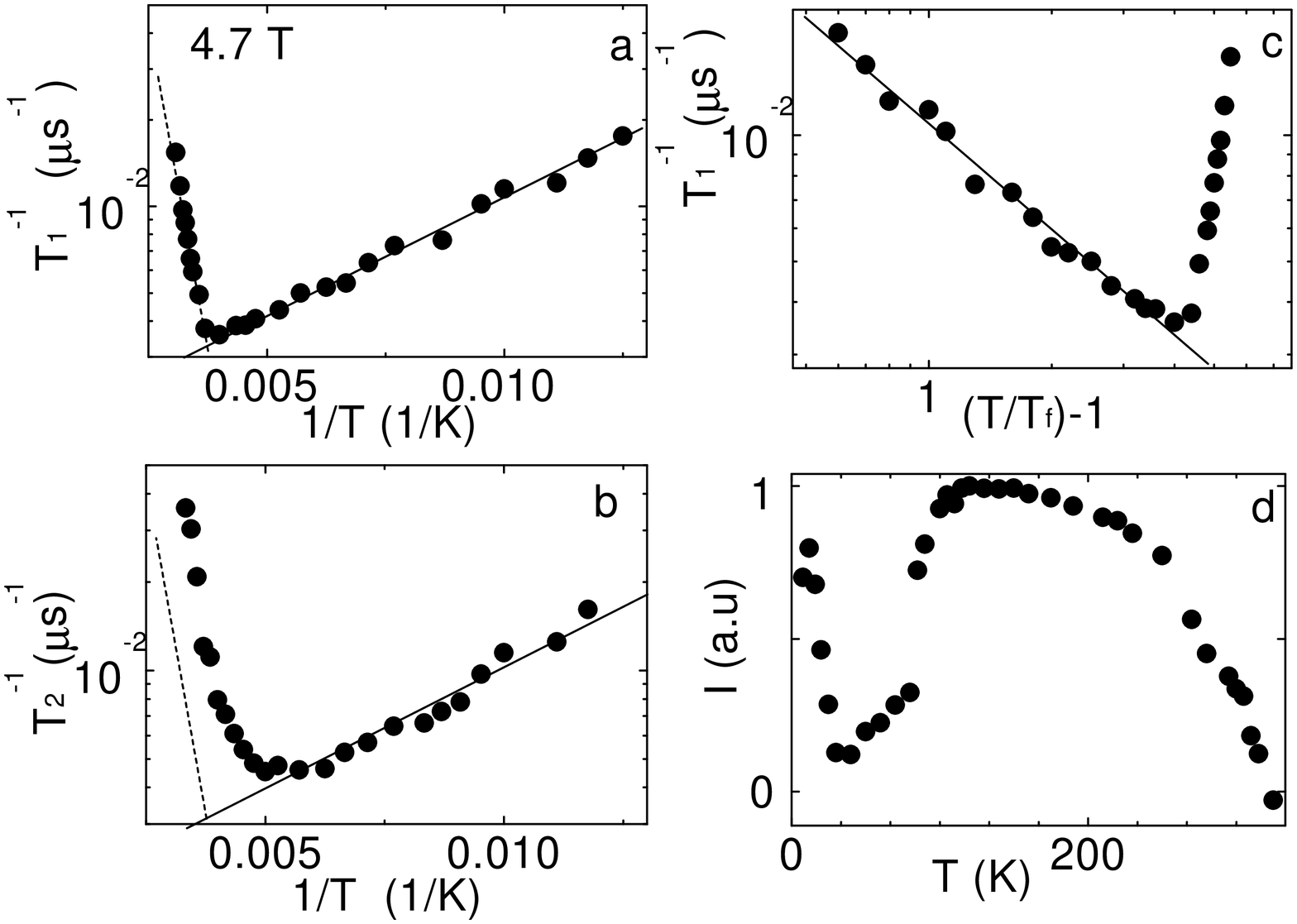,height=6.5cm,width=6.5cm,angle=0}
\end{center}
\caption{(a) $T_1^{-1}$ in 4.7~T on a semilog scale. Straight lines are fits
to activated behavior (b) $T_2^{-1}$ in 4.7~T, scales and lines are the same
as in (a). (c) $T_1^{-1}$ $vs.$ $(T/T_f)-1$ on a log scale.  The straight
line is a fit to the power law expression in the text. (d) Normalized
intensity plot. Above 300~K not all La-nuclei are observed.  Around 50~K a
similar intensity loss occurs.}
\label{f4}
\end{figure}
For $T \downarrow 200$~K a shoulder on the low frequency side of the
resonance peak develops.  For 300~K $>T>$ 230~K the width at 4.7~T is about
two times larger than at 9.4~T.  Between 200~K $>T>$ 150~K the main line
becomes asymmetric being steeper on the low and smoother on the high
frequency side (opposite to the behavior above 230~K) and can be decomposed
into two lines, one centered around the ``250~K'' position (referred to as
line A) and one shifted to higher frequencies (line B), see Fig.~3a.  The
intensity ratio of B to A depends on the cooling history, being larger when
cooling proceeds slower. In Fig.~2 we show typical profiles obtained after
slowly cooling down.  In the decomposition the linewidth of line A was kept
larger than or equal to 250 kHz -- the value at 250~K in 9.4~T.  Down to
130~K the width of line A stayed at this mimimum value, while the intensity
ratio of B to A saturated at 2.  The extra contributions to the widths (and
shifts) in 4.7~T are smaller than in 9.4~T, indicative for a paramagnetic
contribution to the line, in accordance with previous susceptibility
measurements that showed the linewidths to be proportional to the
susceptibility \cite{Bernal97}. This result is well explained by the
development of an internal field in the perpendicular plane, of which the
contribution along the field direction goes linear with the applied magnetic
field.  In the fit to the total linewidth below 200~K, see dashed line in
Fig.~3, the A-linewidth is fixed to 250 kHz (the 250~K width), while the
width of B is described by the expression also used e.g.  by Suh et
al.\cite{Suh98} for the splitting of the doublet seen in NQR:
$\Delta(T)=\Delta_0(1-T/T_N)^{\beta}$, with $T_N \approx $ 200~K and $\beta
\approx 0.5$.  This phenomenological expression is consistent with the growth
of the expectation value of the magnetization. Finally we note that the
linewidth measured at 23.25 MHz at 2~K equals 0.7~T or 4.5 MHz (FWHM), which
is about 5 times larger than at 80~K ($\sim 1$~MHz), and is about an order of
magnitude higher than the calculated effect of the dipolar field (about
0.07~T).  \\

The spin-lattice relaxation rates ($T_1^{-1}$), see Fig.~4, were
measured by a $\pi$--$\tau$--$\pi$ pulse sequence or with a short pulse train
of saturation pulses followed by the usual $\pi/2$--$\tau$--$\pi$ sequence
and are analyzed with the multiexponential expression of Narath
\cite{Narath67}.  The effective relaxation rates related to initial slopes
are plotted in our figures. In case of magnetic relaxation $T_1$ is related
to the fundamental magnetic transition probability $W_M$ via $T_1^{-1}
\approx 46 W_M$.  Above 250~K the relaxation process depends exponentially on
$1/T$.  Around 250~K the slope in Fig.~4 changes in sign and value,
characteristic for a new relaxation process.  In 4.7~T for 230~K $>T>$ 130~K,
$T_1^{-1}$(A) is lower than $T_1^{-1}$(B) with a maximum difference of a
factor 1.5 around 150~K.  In Fig.~4d the normalized NMR intensity (i.e.
corrected for $1/T$, and with proper account for the linewidth and $T_2$)
shows the loss of signal between 60~K and 20~K and above 300~K
\cite{Bernal97}. Below 30~K the spin lattice relaxation rate slows down and
becomes constant below 10~K (not shown). $T_2$ data (Fig.~4b) almost coincide
with $T_1$ below 190~K, while above this temperature the spin dephasing rates
are higher.\\
Above 230~K the relaxation is primarily due to fluctuations of the
$^{139}$La EFG by charge fluctuations (quadrupolar). The relaxation
probability $W_Q$ ($T_1^{-1}$ is of the order of $W_Q$, but depends on the
levels involved in the transition) due the oxygen or charge diffusion can be
written as: $W_Q=A [\tau/(1+\omega^2\tau^2)]$, where A depends on crystal and
ionic parameters \cite{Cohen57}. The correlation time $\tau$ is determined by
the thermally activated hopping process $\tau=\tau_0 \exp(E_a/T)$. From the
Arrhenius behavior we find an activation energy $E_a$ of $3\cdot10^3$~K.
Also the  angular and field dependence of the linewidth are well described by
a quadrupolar interaction. At the highest temperatures the width of the
resonance line is determined by the very short spin dephasing time
(homogeneous broadening).
The increasing asymmetric width and increasing intensity of the
resonance curve with lowering $T$ is explained by an increase in the number
of visible La-sites.\\
For 230~K $>T>$ 80~K there are two centers of gravity in the line
profile.  The increasing relaxation rates with decreasing temperature of line
A (which equals those of line B below 130~K) are typical for a slowing down
of magnetic fluctuations.  Fits can be made with an activated process
$T_1^{- 1} \propto \exp (\Delta E_a/T)$ with $\Delta E_a = 180$~K (Fig.~4a)
or a power law dependence $T_1^{-1} \propto [(T-T_f)/T_f]^{\alpha}$ with
$\alpha \sim 1$ (Fig.~4c) and the spin freezing temperature $T_f$ is about
50~K \cite{note0}.  The $T$ dependences are reminiscent to those seen in
La$_2$Cu$_{1-x}$Li$_x$O$_4$ \cite{Suh98} or Sr doped La$_2$CuO$_4$
\cite{Chou93} above the spin freezing temperature \cite{note0}.  The magnetic
character of the relaxation mechanism below 200~K was confirmed by relaxation
measurements in the (-3/2)--(-1/2) satellite, which gave the same fundamental
transition probability. The decrease of $T_1^{-1}$ below 50~K is also as
expected in such a freezing scenario.  The constant relaxation rates at the
lowest temperatures are most likely due to diffusion: the spin packet excited
by the rf pulses can relax faster to its non excited neighbors than to the
lattice.

Let us now turn to the interpretation of the data. Consistent with the
neutron data \cite{Lee97}, we find three temperature regimes:\\
(i) $250 > T > 190$~K: starting at 250~K a line broadening is observed
which is clearly related to the charge sector (Fig.~3). The relaxation
data (Fig.~4) reveal a strong, charge related relaxation process at
$T > 230$~K. This could be related to the stripe charge ordering
transition. However, since similar values for the activation energy
have been found in La$_2$CuO$_4$ \cite{Rubini94},  it seems more likely
that this relaxation is due  to mobile interstitial  oxygens.
These O-interstitials tilt the neigboring NiO$_6$ octahedra and
influence the electric field gradients at the nearby La-sites. In this way
they are the source for an extra line (at 56.50 MHz in Fig.~2). Above 250~K
the oxygen motion becomes such an effective spin dephasing channel, that the
additonal line disappears.  Unfortunately, it appears that this oxygen
diffusion corrupts the NMR response for $T > 230$ K which makes it impossible
to study the charge ordering transition region directly via the nuclear
relaxation rates.\\
(ii) From the line shape and spin lattice relaxation rates it follows that
around 230~K oxygen diffusion has stopped and charge order is established.
Below 200~K two inequivalent nuclei are seen (line A and B, Fig.~2). The
negligible shift of line A compared to the large positive shift of line B is
consistent with an assigment of line A to non-magnetic, site-centered and
well localized domain walls  and line B to magnetic domains. The intensity
ratio of two to one for $T \leq 150$~K (established when the domain magnetic
order is fully developed, see inset (b) in Fig.~3) confirms such an
assigment.  \\
(iii) The temperature dependence of the width of the B (domain) line
(Fig.~2) is consistent with the presence of static 2D spin order at
temperatures $T < 190$~K. This temperature is consistent with both the
neutron data \cite{Lee97} and the results found in zero field $\mu$SR in
La$_{2-x}$Sr$_x$NiO$_4$ for $x$=0.33 \cite{Chow96}, where a magnetic
transition temperature of 180~K was reported.  Also Raman data
\cite{Blumberg98,Yamamoto98} are in favor of (almost) static two dimensional
charge/spin correlations below 190~K. At the same time, the `A-line',
ascribed to the domain walls, does not show noticable changes down to
temperatures $\simeq$ 50~K indicating that the domain wall spins remain
disordered. Note that above 190~K an extra contribution speeds up the spin
dephasing ($T_2^{-1}$), whithout affecting $T_1^{-1}$.\\
(iv) A transition to a completely frozen spin state occurs at 50~K, which not
only locks in the 2D domain spin system, but also the spin system confined on
the domain walls.  This follows not only from the evolution of the line
widths (Fig.~2), but especially from the intensity loss around 50~K (Fig.~4d)
and the $1/T_1$ data (Fig.~4c), suggesting  the spin-lattice relaxation to be
dominated by the precursor fluctuations of this transition.\\

It appears that the spin-charge coupled ordering dynamics exhibits a
complexity which exceeds by far the theoretical
expectations \cite{Zachar97,vanDuin98}. The neutron work \cite{Lee97} already
pointed at the importance of two dimensional melting physics \cite{Nelson79}
in the charge sector. The hexatic- or glassy charge ordered state was found
to terminate at the $190$~K transition, coinciding with the onset of
domain-magnetic order. Although the evolution of the magnetic order
parameter \cite{vanDuin98} shows the signature of a second order
behavior(Fig.~3), the abundance ratio of the two distiguishable sites
exhibits a much more discontinuous behavior (inset (b) in Fig.~3). Since this
quantity is much less sensitive to long wavelength magnetic
fluctuations \cite{vanDuin98}, it might reflect the lingering first order
behavior expected for a charge-spin coupled system \cite{Zachar97}.\\
In addition, the observations show that although the domain spin system
is ordered for $T < 190$~K, the $S=1/2$ spins on the domain walls remain in a
disordered state: the overall spin system decouples into an ordered state and
a fluctuating subsystem with a 1D appearance. Prelimenary $1/T_1$
results\cite{note1} in fact indicate that the latter exhibits a one
dimensional spin diffusional dynamics in the temperature interval $50 < T <
190$~K, yielding  support for the general notion that fluctuations can
`dynamically' reduce the effective dimensionality of the system, as put
forward in the context of cuprate physics \cite{Emery97}.\\
Finally, the most puzzling aspect is related to the role of quenched
disorder. It seems established that Sr doped samples are more dirty than the
O doped system we have been studying. At the one hand, we find a close
agreement with the neutron study by Lee and Cheong \cite{Lee97} who dealt
with a Sr sample. On the other hand, although the NMR data of Yoshinary {\em
et al.} \cite{Yoshinary98} on the Sr doped nickelate have many features in
common with our findings, there is a striking difference which escapes our
present understanding: although two inequivalent sites are found in the Sr
doped case, their relative abundance shows a very different temperature
depence which is not consistent with site ordered stripes.

We gratefully acknowledge fruitful discussions with S. Mukhin, D.E.
MacLaughlin, Y.  Yoshinari and P.C. Hammel.  One batch of the single
crystals was prepared by Y.M.  Mukovskii at the Steel and Alloys
Institute
in Moscow.

\end{multicols}{2}


\begin{thebibliography}{99}
\bibitem{note0} We observed a dependence of $T_1$ on waiting time
(e.g. at 100~K in 4.7~T the fundamental rate $W$ changed from $580 {\rm s}^{-
1}$ to $360 {\rm s}^{-1}$ by repeated cycling to 200~K and back
during two days), which likely results from magnetic and charge
freezing.
\end{thebibliography}
\end{document}